# Non-selective evaporation mechanism of binary aerosol generating agent on porous atomizer and its experimental verification

XIE Guoyong[1], DU Wen[*,1], WANG Zhiguo[1], CHEN Jingbo[1], SUN Zhiwei[1], LIU Liyan[2], LI Bingbing[2]

1. Technology Center, China Tobacco Hunan Industrial Co., Ltd., Changsha 410007, China
2. College of Chemical Engineering, Tianjin University, Tianjin 300350, China
3. Chemical Engineering Department of Ren'ai College, Tianjin University, Tianjin 301636, China

**Abstract:** To explain the similarity of the compositon of e-cigarette aerosol and its e-liquid source despite of their different boiling points of the components, non-selective evaporation mechanism of binary aerosol generating agent (AGA) was studied based on the phase diagram and capillary evaporation theory. The formation conditions and process of non-selective evaporation were verified by evaporation temperature measurement, airflow heating evaporation experiment and capillary evaporation progress observation. 1) The non-selective evaporation mechanism was deduced as: Because of the obstruction of liquid convection inside the capillary, an evaporation layer whose composition was different from the main liquid phase was gradually formed near the evaporation surface. When the composition of the gas components evaporated from the evaporation layer was consistent with the composition of the main components of the liquid phase transferred into the evaporation layer, a continuous and stable non-selective evaporation was formed.； 2) The evaporation temperature T was measured. The results showed that T deviated from the bubble point temperature $T_b$ obviously and was much closer to the dew point temperature $T_c$ of the gas phase with the same composition ratio of the main liquid phase. This implied that there was a stable evaporation layer with a composition different from that of the main liquid phase; 3) The results of air flow and porous surface heating evaporation experiments showed that under the surface heating conditions, the evaporation of water/methanol, water/propylene glycol and propylene glycol/glycerol binary systems significantly deviated from the ideal selective evaporation, showing an obvious feature of non-selective evaporation. When the liquid surface was covered with cotton fiber felt porous material, the evaporation of binary systems deviated from the selective evaporation furthermore. The important influence of "surface heating" and "mass transfer resistance" on non-selective evaporation was verified; 4) The observation results of the two-stage single capillary evaporation experimental platform showed that the evaporation rate of the water/methanol binary system had an obvious inflection point from fast to slow, indicating that the composition of the evaporation layer changed from consistent with the liquid main body in the initial stage to inconsistent with the liquid main body in the stable non-selective evaporation stage; 5) The particle tracing experiment was applied to observe the overall liquid flow in the capillary tube. The observation results showed that the overall liquid flow in the capillary tube was one-way, there was no large-scale convective mass transfer, and there was a micro convection area near the capillary meniscus, which further verified the existence of the "evaporation layer".

**Keywords:** Binary atomizing agent; Non-selective evaporation; Atomizing mechanism; Surface heating; Mass transfer resistance; Capillary evaporation

## 1 Introduction

The atomization process in electrically heated e-cigarettes involves two stages: the evaporation of e-liquid and its subsequent condensation. Among these, the evaporation of e-liquid from porous medium atomizers is a critical factor in aerosol formation and smoking quality [1–2]. The atomization behavior and mechanisms of e-liquid in porous media are central issues in the development of e-cigarette products.

In recent years, researchers have gained significant insights into the patterns and influencing factors of e-cigarette aerosol generation through experimental measurements and numerical simulations. Talih et al. [3–4] established energy balance equations to predict e-liquid evaporation rates, temperature distributions near heating coils, and changes in aerosol composition. Duan Yuanxing et al. [5] demonstrated that puff duration and puff volume are key factors affecting nicotine release in e-cigarette aerosols, with increased heating power and propylene glycol (PG) content in e-liquids enhancing nicotine release. Pourchez et al. [6] designed experiments to capture e-cigarette aerosols and found that higher heating power and glycerol (VG) content in e-liquids effectively increased total particulate matter (TPM) and average aerosol particle size. In these models, the evaporation of PG and VG follows the phase transition rules of non-azeotropic fluids, meaning the evaporation process is "selective" [7], with low-boiling-point components preferentially evaporating into the gas phase at higher proportions.

However, many experimental data [3, 8] indicate that in e-cigarette atomizer, the ratio of PG to VG in the aerosol closely matches their ratio in the e-liquid, rather than exhibiting "selective" evaporation based on boiling point differences. Previous temperature measurements of e-liquid atomization by our research group [9] also revealed that during e-cigarette atomization, PG-VG mixtures of varying proportions correspond to a stable atomization temperature, which lies between the boiling points of PG and VG and increases with higher VG mass fractions. These results suggest that the atomization of PG-VG mixtures in e-cigarette atomizers is "non-selective".

To date, no comprehensive theory has been proposed to explain the non-selective atomization characteristics of e-cigarette atomizers successfully. One hypothesis is that the atomization process is explosive—part of the e-liquid is instantaneously vaporized, causing rapid volume expansion and generating shock waves that break nearby liquid into droplets, which then enter the aerosol. However, this hypothesis was disproven by NaCl addition experiments. Talih et al. [4] demonstrated that over 95% of the aerosol is produced through an evaporation-condensation mechanism.

To uncover the mechanisms behind the "non-selective evaporation" phenomenon in multi-component atomizing agents and explore the key factors influencing their evaporation and aerosol release in porous media, this study applies phase diagrams and capillary evaporation theory to derive the "non-selective evaporation" mechanism for binary systems. The mechanism was validated through evaporation temperature measurements, airflow heating evaporation experiments, and the observations of capillary evaporation processes.

## 2 Materials and Methods
### 2.1 Materials, Reagents, and Instruments
Propylene glycol (PG), glycerol (VG), and methanol (AR, Sinopharm Chemical Reagent Co., Ltd.) were used.

The LB100 cigarette combustion temperature detection system (sampling frequency: 20 Hz, Beijing Liboxin Technology Co., Ltd.) was equipped with ultra-fine K-type thermocouples (Omega, USA) with a diameter of 0.254 mm, temperature detection accuracy of 0.3 ℃, and response time of 0.1 s. Additional instruments included a FASTCAM Mini AX200 CCD camera (Photron, Japan), an IX73 inverted microscope (Olympus, Japan), and an S3050-15K ultraviolet light source (Zhuhai Tianchuang Instrument Co., Ltd.).

## 2.2 Methods
### 2.2.1 Calculation of Liquid-Gas Equilibrium Phase Diagrams for Binary Systems

The vapor pressure of pure substances at different temperatures was calculated using the Antoine equation [10] [Equation (1)]:

$$\lg P = A + \frac{B}{T+C} + 5 \tag{1}$$

where $P$ is the vapor pressure (Pa), A, B, and C are Antoine constants (dimensionless), and $T$ is the temperature (K).

For a binary liquid-gas equilibrium system, let $P_1$ and $P_2$ be the vapor pressures of pure components 1 and 2, respectively. If the mole fraction of component 1 in the liquid phase is $R_1$, the partial pressure of component 1 in the gas phase is $P_1 R_1$, and that of component 2 is $P_2(1-R_1)$. At the bubble point temperature $T_b$, the sum of the component vapor pressures equals the ambient atmospheric pressure $P_0$, as shown in Equation (2):

$$P_0 = P_1 R_1 + P_2(1 - R_1) \tag{2}$$

Using Equations (1) and (2), the bubble point temperature $T_b$ for binary systems with different liquid-phase compositions can be calculated.

At the bubble point temperature, the mole fraction of component 1 in the gas phase $G1$ is calculated using Equation (3):

$$G_1 = \frac{P_1 R_1}{P_0} \tag{3}$$

The liquid-gas equilibrium phase diagram for the binary system can then be plotted using $T_b$, $R_1$ and $G_1$.

Phase diagram calculations and plotting were performed in the Matlab 2010(b) environment.

### 2.2.2 Calculation of Liquid Component Fraction Changes in Ideal Selective Evaporation Processes

Consider an "ideal selective evaporation" process for a binary system under constant temperature with sufficient liquid-phase convection. Let the initial total liquid-phase mole quantity $M_0$ be 1 mol. As evaporation proceeds, the liquid-phase quantity $M$ decreases. At any moment, the change in liquid-phase quantity d$M$ and the change in mole fraction of component 1 d$R_1$ are related as follows:

$$(M + dM)(R_1 + dR_1) - M R_1 = \frac{G_1}{G_1 + G_2} dM \tag{4}$$

where $MR_1$ is the content of component 1 in the liquid phase before evaporation, $(M+dM)(R_1+dR_1)$ is the content of component 1 in the remaining liquid phase after evaporating d$M$, and $G_1/(G_1+G_2)$d$M$ is the content of component 1 in the evaporated gas phase (mol). $G_1$ and $G_2$ are calculated using Equation (3). Solving this ordinary differential equation yields the $R_1$~$M$ curve for the "ideal selective evaporation" process.

The relationship between liquid-phase quantity $M$ (mol) and liquid-phase volume $V$ (m³) is given by:

$$V = MR_1/\rho_{m1} + M(1 - R_1)/\rho_{m2} \qquad (5)$$

where $\rho_{m1}$ and $\rho_{m2}$ are the densities of components 1 and 2 (mol/m³), respectively. Letting the initial liquid-phase volume be $V_0$, the $R_1$~$M$ curve can be transformed into an $R_1$~$V/V_0$ curve, which is independent of the initial liquid volume and depends only on the properties and initial proportions of the components.

### 2.2.3 Atomization Temperature Measurement

As shown in Figure 1, the atomizer was opened, and a thermocouple was placed close to the interface between the heating coil and the absorbent cotton. The atomizer's heating coil had a resistance of 0.54 Ω and supply power was set to 15 W. Temperature was measured using a 0.254 mm ultra-fine K-type thermocouple with a data acquisition frequency of 20 Hz.

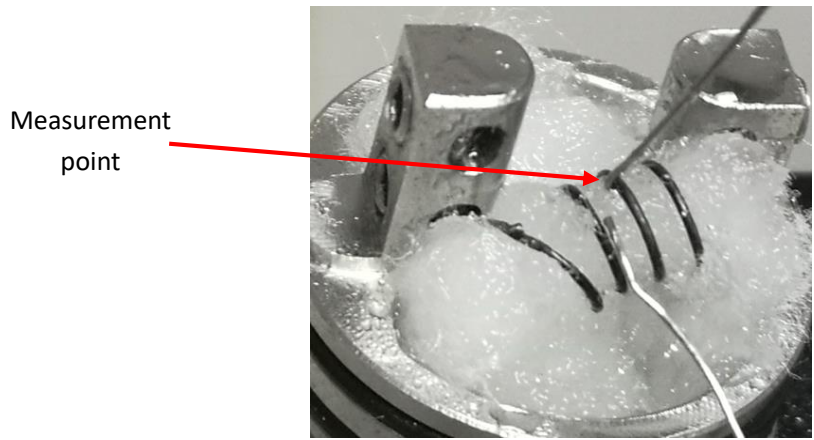

**Figure 1. Atomization Temperature Measurement Aparatus**

### 2.2.4 Surface Heating and Liquid-Phase Mass Transfer Resistance Experiments

To study the effect of surface heating on the evaporation selectivity of binary mixtures, an airflow surface heating experimental setup was constructed (Figure 2). Dry air was heated to a set temperature and introduced into an evaporation flask containing a binary mixture. The liquid surface was heated by the airflow, and the evaporated gas was discharged through an outlet. $G_1$, $G_2$, $F_1$ and $F_2$ represent the mole fractions of components 1 and 2 in the gas and liquid phases, respectively. A sampling valve at the bottom of the flask allowed periodic sampling to analyze the mole fractions of the liquid components, while changes in the liquid level were recorded.

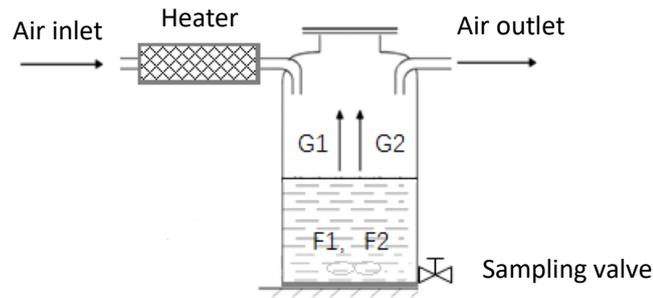

Figure 2. Airflow Surface Heating Evaporation Experimental Apparatus

To investigate the effect of liquid-phase mass transfer resistance on evaporation selectivity, a porous medium surface heating evaporation setup was designed based on the airflow heating setup (Figure 3). A cotton fiber felt porous medium (Figure 3b) was placed on a wooden floater (Figure 3c), with the lower part of the felt immersed in the liquid. The liquid was draw to the upper surface of the felt by capillary force.

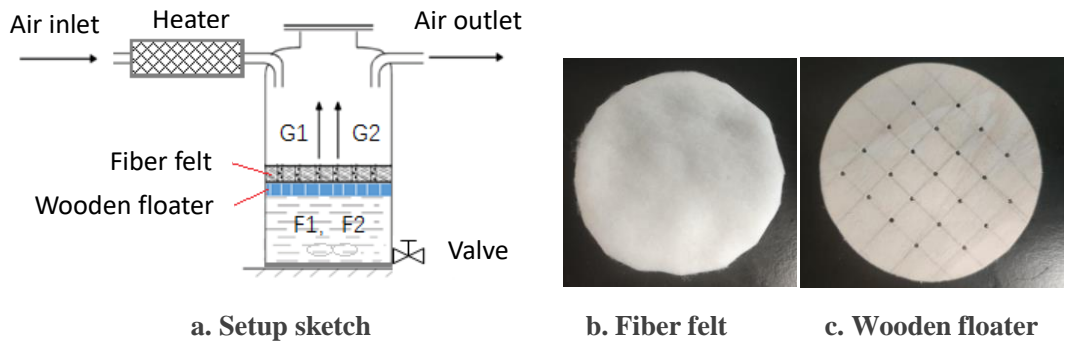

a. Setup sketch　　　　　b. Fiber felt　　　c. Wooden floater
Figure 3. Porous Medium Surface Heating Evaporation Experimental Setup

Binary mixtures of water/methanol, water/PG, and PG/VG were prepared with a mole ratio of 5:5. A total of 200 mL of each mixture was added to the evaporation flask, with the initial volume recorded as $V_0$. Hot air temperatures were set to 70, 120 and 180 °C, with an airflow rate of 7.5 L/min. During evaporation, liquid samples were taken from the flask to analyze the mole fractions ($F$) of the remaining components, while the remaining liquid volume ($V$) was recorded. Sampling continued until the liquid was fully evaporated. The change in $F$ with $V/V_0$ was used to evaluate the selectivity of evaporation. Smaller changes in $F$ indicated stronger non-selective evaporation.

The evaporation selectivity of binary systems was compared for free surface heating without porous media, with 4 mm and 10 mm thick cotton fiber felts.

### 2.2.5 Capillary Evaporation Rate Observation

A horizontal combined capillary setup was used to study the evaporation process (Figure 4). The setup consisted of a smaller-diameter heating evaporation section and a larger-diameter liquid injection observation section, connected by a sealed rubber tube. The difference in capillary pressure between the two sections kept the meniscus in the heating section stationary, while the meniscus in the observation section moved toward the heating section to replenish evaporated liquid, maintaining an evaporation-replenishment balance. The evaporation rate was calculated by the movement of the meniscus in the observation section.

This dual-section capillary setup ensured a constant evaporation surface position and surrounding environment, avoiding issues in single-capillary experiments where the liquid surface detached from the tube opening, affecting gas-liquid equilibrium in the narrow space.

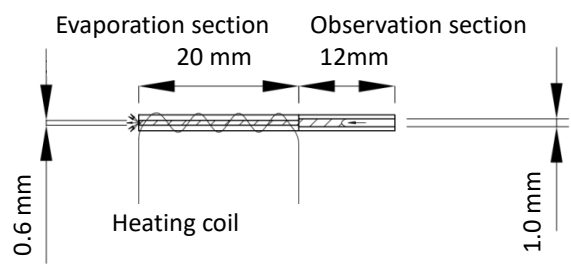

Figure 4. Horizontal Combined Capillary for Evaporation Experiments

### 2.2.6 Microscopic Observation of Liquid Flow in Capillary Evaporation
To observe liquid mass transfer characteristics during capillary evaporation, a microscopic observation setup was designed (Figure 5). The setup included a microscope, CCD camera, ultraviolet light source, and capillary evaporation device. Fluorescent polystyrene microspheres (5 μm) were used as tracer particles to observe fluid flow patterns during evaporation. The capillary (inner diameter: 0.6 mm) was horizontally fixed on the microscope stage using high-temperature tape. Methanol was used as the evaporating fluid, and tracer particles were introduced into the capillary using a glass rod. The motion of the particles was recorded using the CCD camera focused on the microscope objective.

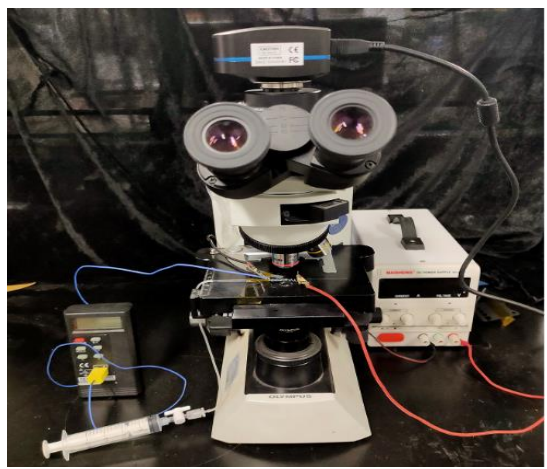

Figure 5. Microscopic Observation Devices for Capillary Evaporation with Tracer Particles

### 3 Results and Discussion
### 3.1 Selective Evaporation of Binary Mixtures
The liquid-gas equilibrium phase diagram for the glycerol (VG)-propylene glycol (PG) binary system was plotted using the Antoine parameters listed in Table 1 and the method described in Section 2.2.1. As shown in Figure 6, line **a** represents the bubble points under different liquid-phase compositions. When the mass fraction of VG in the liquid phase ranges from 0 to 1, the bubble point temperature varies between 460 and 560 K, increasing with higher VG mass fractions. When the VG mass fraction is 0, the system contains only PG, and the bubble point temperature is the boiling point of PG (460 K). When the VG mass fraction is 1, the system contains only VG, and the bubble point temperature is the boiling point of VG (560 K).

**Table 1 Antoine Constants for Glycerol (VG) and Propylene Glycol (PG)** [10]

|    | Boiling point/K | A       | B        | C        |
|----|-----------------|---------|----------|----------|
| VG | 560             | 3.93737 | 1411.531 | −200.566 |
| PG | 460             | 6.07936 | 2692.187 | −17.94   |

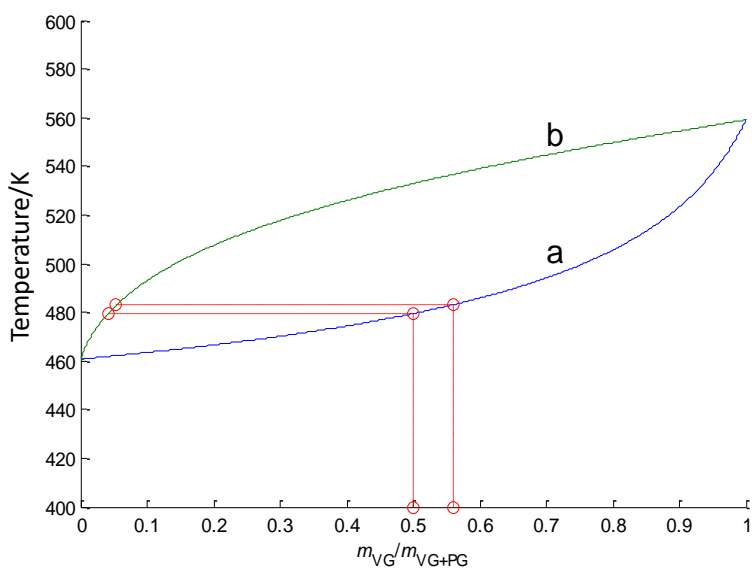

a) Bubble Point Line, b) Dew Point Line
**Figure 6. Liquid-Gas Equilibrium Phase Diagram for the VG-PG Binary System**

In Figure 6, line **b** represents the dew points under different liquid-phase compositions. From this line, the gas-phase fractions at specific bubble point temperatures can be determined. For example, when the VG mass fraction in the liquid phase is 0.5, the bubble point temperature is 479 K. At this temperature, the VG mass fraction in the gas phase is 0.04, and the PG mass fraction is 1 - 0.04 = 0.96. This indicates that the lower-boiling-point component (PG) is preferentially evaporated, which is the fundamental principle of distillation. The significant difference between the gas-phase and liquid-phase composition ratios during evaporation is referred to as "selective evaporation" in this study.

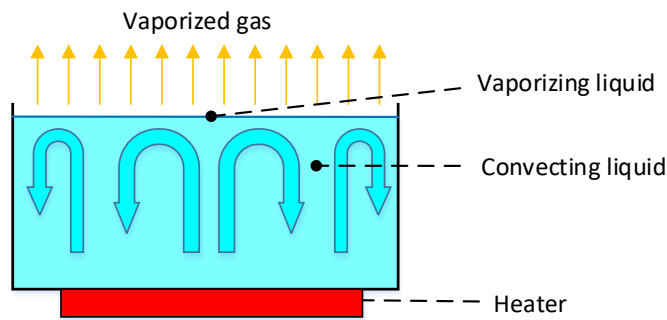

**Figure 7. Schematic Diagram of a Bottom-Heated Selective Evaporation System for the VG-PG Binary Mixture**

As shown in Figure 7, the bottom-heated selective evaporation system for the VG-PG mixture consists of a container filled with the mixture. The liquid at the bottom is heated first, creating convection within the container, and evaporation occurs at the liquid-gas interface. According to the phase diagram in Figure 6, when the VG mass fraction in the liquid phase is 0.5, the VG mass fraction in the evaporated gas phase is 0.04, indicating that PG is preferentially evaporated. This changes the composition at the evaporation surface, where the PG mass fraction becomes lower than that in the bulk convection zone. However, this compositional change is transient, as the evaporation surface composition is rapidly updated by liquid-phase convection, ensuring that the evaporating liquid composition remains consistent with the bulk convection zone. Due to the selectivity of evaporation, as evaporation continues, the PG mass fraction in the gas phase remains higher than that in the liquid phase, causing the PG mass fraction in the liquid phase to gradually decrease and the VG mass fraction to increase. This is reflected in the phase diagram (Figure 6), where the bubble point gradually rises and moves upward along the bubble point line until all PG is evaporated, leaving only pure VG with a constant boiling point.

### 3.2 Non-Selective Evaporation in E-Liquid Atomization

Zhang et al. [8] prepared an e-liquid with a mass ratio of VG to PG of 45:35 (VG mass fraction of 0.56 in the VG-PG mixture) and measured the release of VG and PG in the aerosol over 60 puffs. The results showed that under heating powers of 3.9–13.6 W, the VG mass fraction in the aerosol ranged from 0.54 to 0.61, significantly higher than the theoretical value of 0.05 predicted by the equilibrium phase diagram (Figure 6) and closely matching the composition of the e-liquid. This indicates that e-liquid atomization significantly deviates from "selective evaporation."

When the gas-phase composition closely matches the liquid-phase composition and is independent of the boiling points and saturated vapor pressures of the components, this phenomenon is referred to as "non-selective evaporation" in this study. Extensive testing has shown that e-cigarette aerosols exhibit clear "non-selective evaporation" characteristics, regardless of whether porous ceramic atomizers or cotton-wick heating coil atomizers are used.

### 3.3 Mechanism of Non-Selective Evaporation in Binary Systems

E-cigarette atomizer cores primarily consist of porous ceramic cores and cotton wicks, both of which are porous materials with capillary-like structures. The "non-selective evaporation" phenomenon of e-liquids is closely related to these capillary structures. Figure 8 shows a schematic of a single capillary in a porous ceramic atomizer, which is composed of numerous capillary evaporators. Figure 9 illustrates a cotton-wick atomizer and its local structure, where the gaps between cotton fibers form porous liquid-conducting channels. These channels exhibit capillary effects similar to those in Figure 8, with the key difference being that capillary action occurs on the outer surfaces of the fibers rather than the inner surfaces of capillaries.

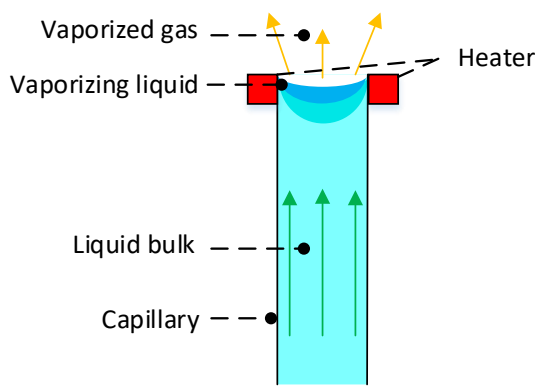

Figure 8. Schematic of a Single Capillary in a Porous Ceramic Atomizer

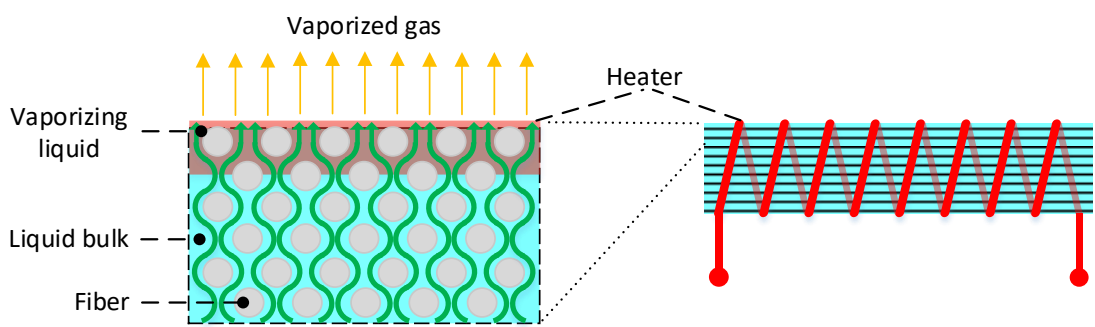

Figure 9. Schematic of a Cotton-Wick Atomizer

Compared to the selective evaporation system shown in Figure 7, e-cigarette atomizers (Figures 8 and 9) have two significant differences: (1) The heat source is located near the evaporation surface rather than on the opposite side, preventing the formation of large-scale temperature-density gradients and hindering large-scale convective mass transfer. (2) The driving force for liquid conduction is capillary action, which is unidirectional along the capillary walls, limiting diffusive mass transfer. Capillary evaporation primarily occurs in the thin film region near the channel walls [11–12]. Based on liquid film thickness and detachment pressure, the extended liquid film in the capillary can be divided into three regions along its axis: the adsorption region, where the thin liquid layer is strongly bound by solid-liquid interactions and difficult to evaporate; the thin film region, which is closer to the heat transfer surface and has significantly lower thermal resistance than the intrinsic meniscus region, leading to preferential evaporation; and the evaporation region, where changes in liquid composition cause surface tension variations, inducing Marangoni flow [11] and driving fluid flow and mass transfer. However, this mass transfer is confined to a small region and has little effect on the bulk liquid composition in the transport region.

Based on the above analysis, the non-selective atomization mechanism of e-cigarette atomizers can be inferred from the characteristics of capillary evaporation. Using the VG-PG binary system with a VG mass fraction of 0.5 as an example, the evaporation process can be analyzed from the phase diagram (Figure 10):

1) The evaporation region is heated to the bubble point temperature $T_b$, initiating vigorous evaporation. At this stage, the temperature is 479 K, the VG mass fraction in the liquid phase is 0.5, and the VG mass fraction in the gas phase is 0.04.
2) PG in the evaporation region is lost significantly faster than VG. Due to hindered mass transfer, the replenishment of the evaporation region by the transport region occurs unidirectionally. As evaporation continues, the VG mass fraction in the evaporation region increases, and the temperature rises accordingly.
3) When the VG mass fraction in the evaporation region reaches 0.94 and the bubble point temperature rises to 533 K (equivalent to the dew point temperature $T_c$ for a gas-phase VG mass fraction of 0.5), the VG mass fraction in the evaporated gas phase reaches 0.5. At this point, the composition of the gas phase evaporating from the evaporation region matches that of the liquid phase entering the evaporation region, achieving an evaporation-transport equilibrium. The evaporation temperature stabilizes at 533 K, and the VG mass fraction in the evaporation region remains at 0.94. VG and PG continue to evaporate into the gas phase at a mass ratio of 0.5:0.5, while the liquid phase is replenished at the same ratio, maintaining a stable "non-selective evaporation" state.

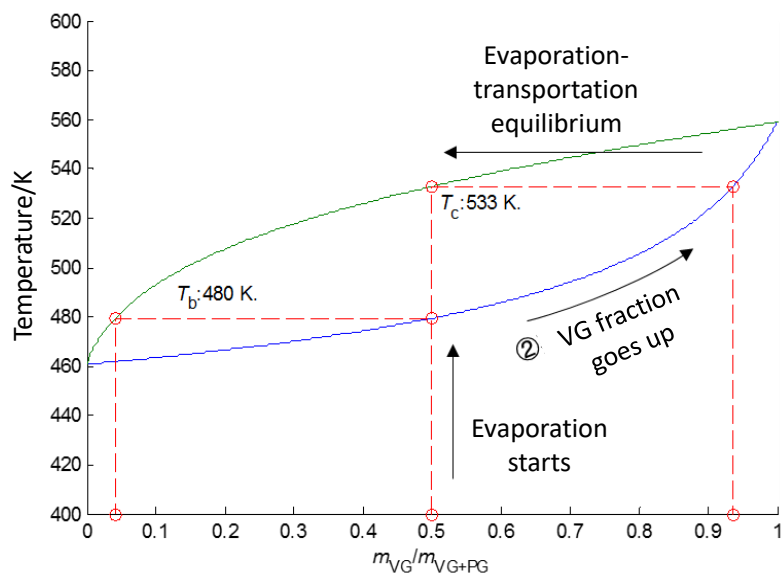

**Figure 10. Schematic of the Non-Selective Evaporation Process in the VG-PG Binary System**

This process can be summarized as follows: Due to hindered mass transfer, an "evaporation layer" and a "transport layer" form during capillary evaporation. The evaporation layer maintains a composition different from that of the transport layer, ensuring that the gas-phase composition evaporating from the evaporation layer matches the liquid-phase composition entering the evaporation layer, resulting in "non-selective evaporation." The key to this mechanism is the existence of an evaporation layer with a composition different from that of the transport layer. Although measuring the composition of such a small region during evaporation is challenging, the evaporation temperature can be measured. As shown in the phase diagram (Figure 10), the evaporation temperature corresponds to the liquid-phase composition. For example, when the VG mass fraction in the liquid phase is 0.5, the evaporation temperature is 479 K; when the VG mass fraction is 0.94, the evaporation temperature is 533 K. Therefore, if a stable evaporation layer exists as described in this mechanism, the measured evaporation temperature should be higher than the bubble point temperature $T_b$ corresponding to the bulk liquid-phase composition and closer to the dew point temperature $T_c$ of the gas phase with the same composition.

### 3.4 Experimental Validation of the Non-Selective Evaporation Mechanism

The conditions for "non-selective evaporation" proposed in the previous discussion are: (1) the heat source is located near the evaporation surface, and (2) convective mass transfer is hindered. This section validates the non-selective evaporation mechanism for binary systems by comparing measured evaporation temperatures with bubble point and dew point temperatures from the phase diagram, analyzing the effects of surface heating and mass transfer resistance on non-selective evaporation through airflow heating experiments, and observing the formation process of non-selective evaporation using a single-capillary experimental platform.

#### 3.4.1 Evaporation Temperature Measurement for Binary Mixtures

A thermocouple was placed close to the heating coil of a cotton-wick atomizer to measure the evaporation temperature during heating for 0–5 s without puffing. Seven VG-PG binary e-liquid samples with VG mass fractions of 0, 0.2, 0.3, 0.4, 0.5, 0.6 and 1.0 were prepared. The temperature measurement results are shown in Figure 11. All samples reached a plateau temperature within 0.8~2.7 s. The plateau temperature could be considered as the stable evaporation temperature. The evaporation temperature for pure PG was 456 K, and for pure VG, it was 559 K, differing from their boiling points (460 K and 560 K, respectively) by only 4 K and 1 K, indicating good accuracy of the measurement method.

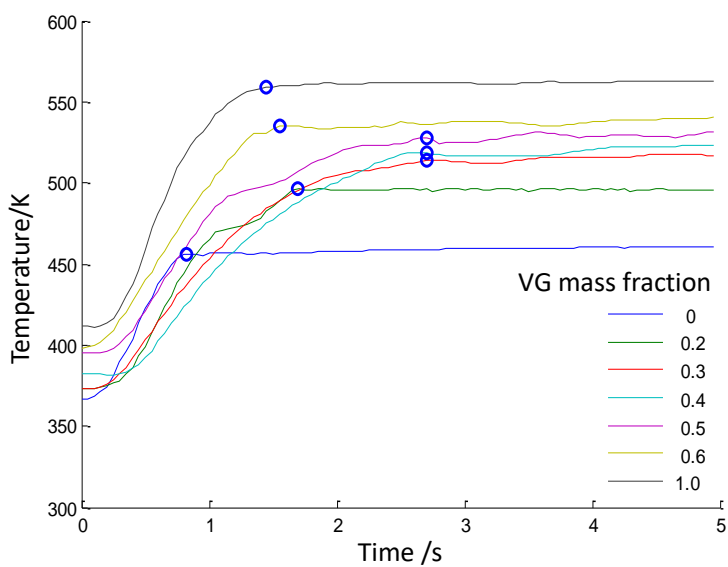

Figure 11. Evaporation Temperature During Atomization in a Cotton-Wick Atomizer

The measured evaporation temperatures $T$ for the seven samples were compared with the bubble point temperatures $T_b$ and dew point temperatures $T_c$ calculated from the phase diagram, as shown in Figure 12. For the five VG-PG binary samples, the measured evaporation temperatures $T$ differed from the bubble point temperatures $T_b$ by 30~49 K but were close to the dew point temperatures $T_c$ for the same composition, differing by only 4~10 K. This confirms the existence of a relatively stable heterogeneous liquid-phase evaporation layer during the evaporation of this binary system. The composition of this evaporation layer differs from that of the bulk liquid phase, causing the gas-phase composition evaporating from the evaporation layer to deviate from that predicted by the bulk liquid-phase composition and instead align more closely with the original liquid-phase composition. Thus, although selective evaporation occurs in the micro-region of the evaporation layer, the overall process exhibits "non-selective evaporation" characteristics.

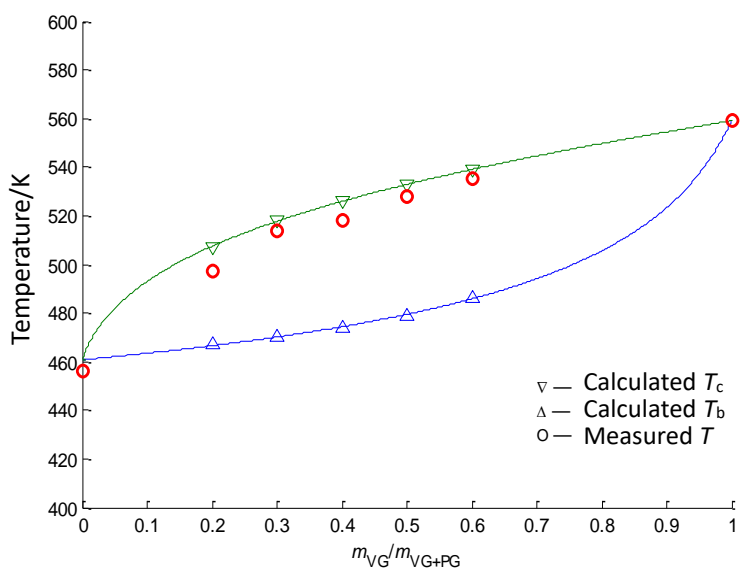

**Figure 12. Comparison of Measured Evaporation Temperatures with Calculated Bubble Point and Dew Point Temperatures**

### 3.4.2 Effects of Surface Heating and Liquid-Phase Mass Transfer Resistance on Evaporation Selectivity

The airflow surface heating experimental setup shown in Figure 2 was used to investigate the effect of surface heating on the evaporation selectivity of binary mixtures. The change in the water mole fraction in the liquid phase during the evaporation of a water-methanol binary mixture under surface heating is shown in Figure 13a. The x-axis represents the ratio of the remaining liquid volume($V$) to the initial liquid volume($V_0$), and the y-axis represents the water mole fraction in the remaining liquid. The dashed line represents the calculated ideal "selective evaporation" trend, while the solid line with "o" markers represents the measured trend under surface heating.

As shown in Figure 13a, compared to ideal "selective evaporation," the surface heating system exhibits significant non-selectivity, with the increase in the water mole fraction being much lower than the ideal trend. This is because, in the water-methanol surface heating system, initial surface evaporation is selective, with methanol being preferentially evaporated. This creates an evaporation layer with a higher water mole fraction at the liquid surface. Since the heat source is located above the surface, the surface liquid is hotter and less dense, preventing strong convection within the liquid. Mass exchange between the evaporation layer and the bulk liquid is driven primarily by concentration gradient diffusion rather than convection, causing the water mole fraction in the evaporation layer to remain higher than that in the bulk liquid. As a result, the water mole fraction in the evaporated gas phase remains higher than in the ideal selective evaporation state, weakening the preferential evaporation of methanol. This leads to more methanol and less water being retained in the remaining liquid as evaporation progresses, and the gas-phase composition becomes closer to the liquid-phase composition, exhibiting "non-selective evaporation" characteristics.

The water-PG (Figure 13b) and PG-VG (Figure 13c) systems under airflow surface heating also show significant deviations from "selective evaporation." This indicates that placing the heat source near the evaporation surface rather than on the opposite side significantly enhances non-selective evaporation.

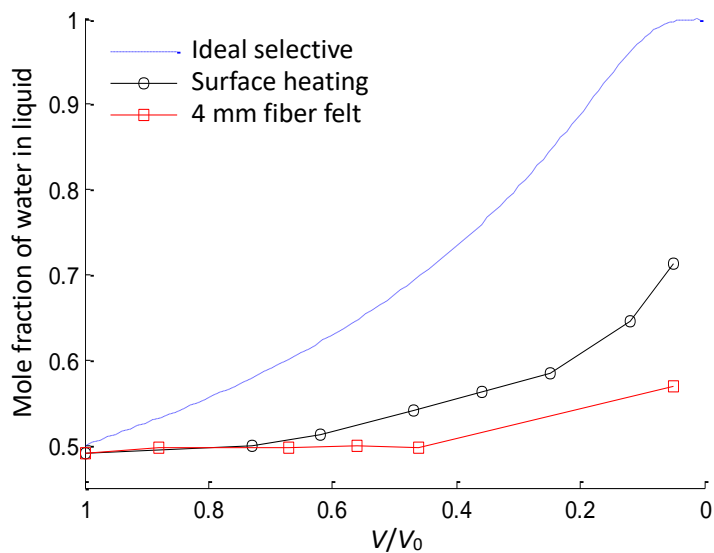

a) **Water-Methanol**

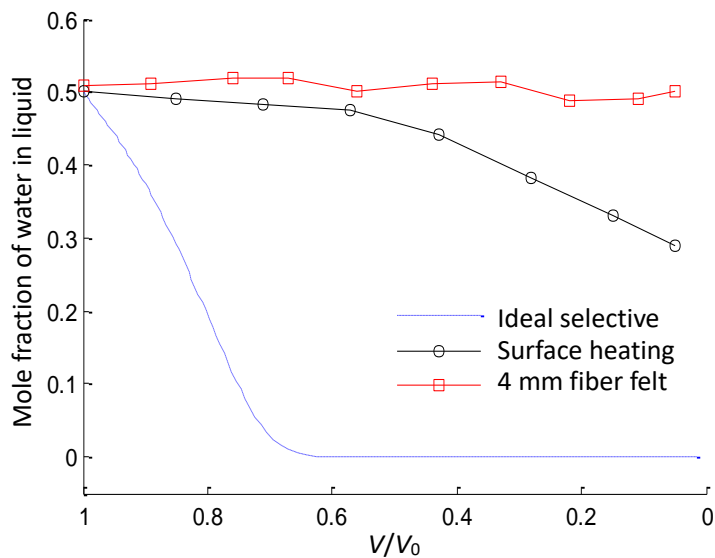

b) **Water-PG**

13surface liquid is hotter and less dense, preventing strong convection within the liquid. Mass exchange between the evaporation layer and the bulk liquid is driven primarily by concentration gradient diffusion rather than convection, causing the water mole fraction in the evaporation layer to remain higher than that in the bulk liquid. As a result, the water mole fraction in the evaporated gas phase remains higher than in the ideal selective evaporation state, weakening the preferential evaporation of methanol. This leads to more methanol and less water being retained in the remaining liquid as evaporation progresses, and the gas-phase composition becomes closer to the liquid-phase composition, exhibiting "non-selective evaporation" characteristics.

The water-PG (Figure 13b) and PG-VG (Figure 13c) systems under airflow surface heating also show significant deviations from "selective evaporation." This indicates that placing the heat source near the evaporation surface rather than on the opposite side significantly enhances non-selective evaporation.

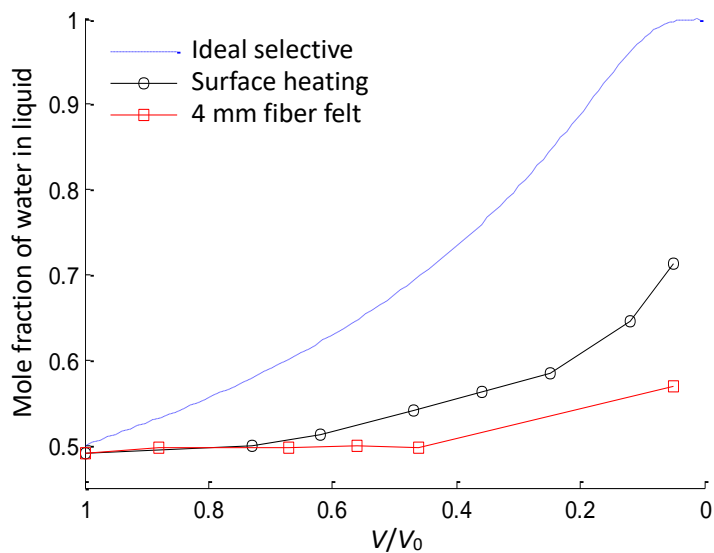

a) **Water-Methanol**

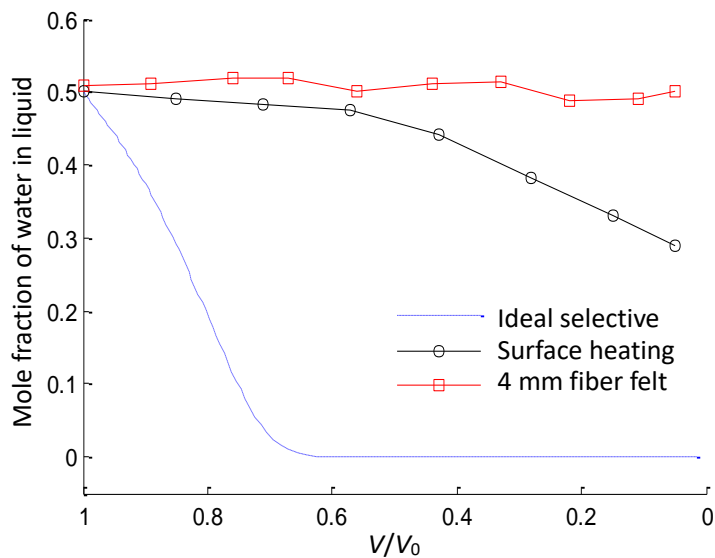

b) **Water-PG**

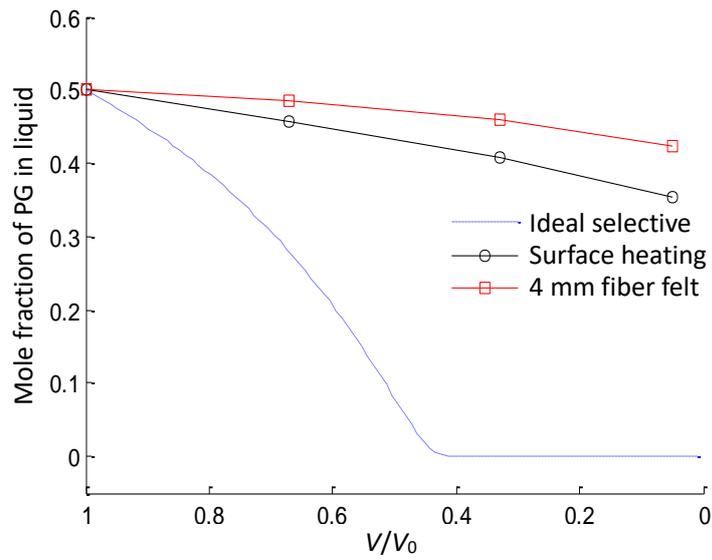

c) PG-VG

**Figure 13. Changes in Liquid-Phase Mole Fraction with Liquid Volume for Binary Systems Under Surface Heating and Ideal "Selective Evaporation"**

Introducing a porous medium at the evaporation surface further hinders mass transfer between the evaporation surface and the bulk liquid, enhancing non-selective evaporation. Using the porous medium surface heating evaporation setup shown in Figure 3, the change in evaporation selectivity before and after adding a 4 mm thick cotton fiber felt was compared.

As shown in Figure 13a, when evaporation occurs on a porous medium (cotton fiber felt), the $F-V/V_0$ curve deviates further from the ideal "selective evaporation" trend. At the end of evaporation on the porous medium, the water mole fraction in the remaining liquid is 0.57, significantly lower than the 0.71 observed for free surface evaporation and the 1.0 for ideal "selective evaporation," and closer to the initial water mole fraction (0.5). This result indicates that evaporation on a porous medium surface is more non-selective than free surface heating evaporation. The presence of the porous medium hinders mass exchange between the surface liquid and the bulk liquid, allowing the surface evaporation layer to maintain a composition different from that of the bulk liquid, with a higher proportion of high-boiling-point components. As a result, the gas-phase composition more closely matches the bulk liquid-phase composition, exhibiting stronger "non-selective evaporation" characteristics.

To further investigate the effect of porous medium mass transfer resistance on evaporation selectivity, the influence of the thickness of the cotton fiber felt was investigated. The changes in $F-V/V_0$ for a water-PG mixture with a mole ratio of 7:3 under free surface evaporation, 4 mm thick porous medium evaporation, and 10 mm thick porous medium evaporation were compared with ideal selective evaporation, as shown in Figure 14. At the start of evaporation, the water mole fraction was 0.7 for all cases. Near the end of evaporation, the water mole fraction was 0.64 for the 10 mm thick porous medium, 0.55 for the 4 mm thick porous medium, and 0.45 for free surface heating. This demonstrates that thicker porous media increase mass transfer resistance, making surface heating evaporation more non-selective.

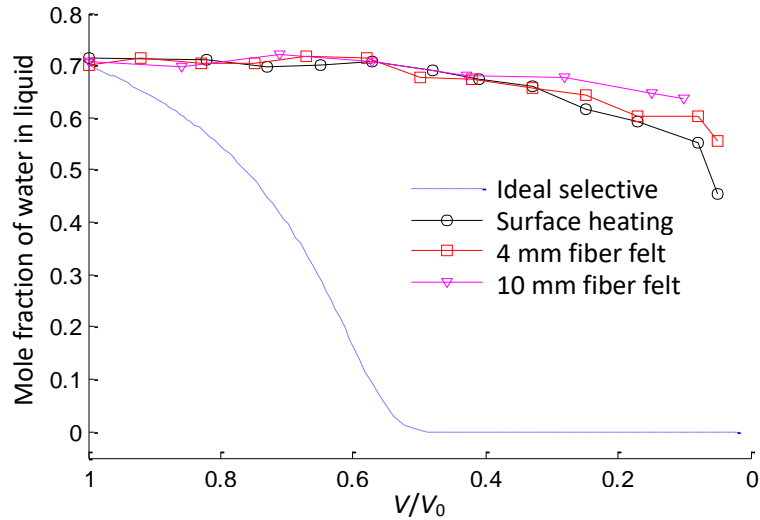

**Figure 14. Changes in Water Mole Fraction with Liquid Volume for Water-PG (7:3 Mole Ratio) Mixture Under Ideal "Selective Evaporation," Free Surface Heating and Porous Medium Surface Heating**

### 3.4.3 Observation of Non-Selective Evaporation Formation

A methanol-water binary mixture with a mole ratio of 3:7 was prepared, and the evaporation process was observed using the horizontal dual-section capillary setup shown in Figure 4. The evaporation temperature was set to 45–75 ℃, and the movement of the liquid surface in the observation tube was recorded in real time using a CCD camera.

The evaporation rates for the 3:7 methanol-water mixture at different temperatures are shown in Figure 15. At 65 ℃, a clear inflection point in the evaporation rate was observed at 250 s, corresponding to the transition from selective to non-selective evaporation. Before the inflection point, the evaporation layer composition was close to the initial system composition, and selective evaporation dominated. As selective evaporation progressed and the bulk liquid replenished the evaporation layer, the low-boiling-point component (methanol) mass fraction in the evaporation layer decreased. Near the inflection point, the methanol mass fraction in the evaporation layer decreased to a level where the gas-phase composition evaporating from the evaporation layer matched the liquid-phase composition entering the evaporation layer, achieving a stable "non-selective evaporation" state. At this point, the methanol mass fraction in the evaporation layer was significantly lower than in the bulk liquid, and the total vapor pressure of methanol and water at 65 ℃ was lower than in the initial evaporation state, causing the evaporation rate to decrease. A similar inflection point was observed at 55 ℃, but it occurred later (400 s) due to the lower temperature. No clear inflection point was observed at 45 ℃, indicating that stable non-selective evaporation was not achieved within the observation period.

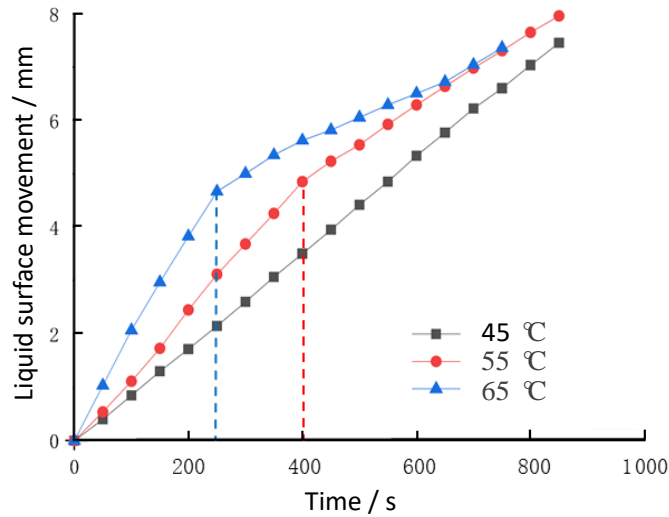

**Figure 15. Evaporation Rates for a 3:7 Methanol-Water Mixture at Different Temperatures**

A similar inflection point was observed in the evaporation of a 1:9 methanol-water mixture, as shown in Figure 16. At 75 ℃, the inflection point occurred at 100 s; at 65 ℃, it occurred at 350 s; and at 55 ℃, no clear inflection point was observed.

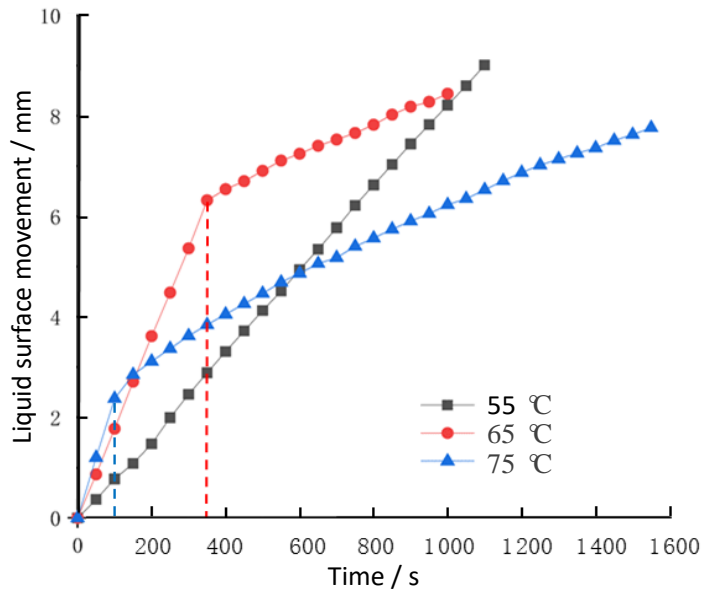

**Figure 16. Evaporation Rates for a 1:9 Methanol-Water Mixture at Different Temperatures**

Comparing the evaporation processes of 5:5, 3:7, and 1:9 methanol-water mixtures at 65 ℃ (Figure 17), the inflection points for the 3:7 and 1:9 mixtures occurred at 250 s and 350 s, respectively. This shows that, at the same evaporation temperature, the time required to achieve non-selective evaporation decreases as the low-boiling-point component mass fraction increases. For the 5:5 mixture, no clear inflection point was observed because the transition from selective to non-selective evaporation occurred within the first observation period (50 s).

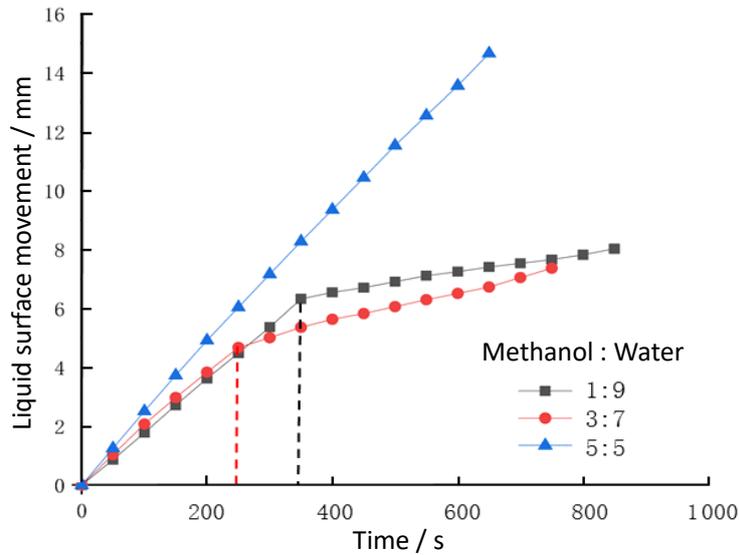

Figure 17. Evaporation Rates for Methanol-Water Mixtures with Different Mole Ratios at 65 ℃

### 3.4.4 Observation of Liquid Flow Direction in Capillary Evaporation

The fluorescent tracer microscopic observation setup shown in Figure 5 was used to observe liquid flow characteristics during capillary evaporation. The original images are shown in Figure 18a, and the instantaneous positions and movement trajectories of the fluorescent particles are shown in Figures 18b and 18c, respectively.

The observations revealed that, upon introducing the fluorescent tracer particles, a large number of particles rapidly flowed upward from the bottom of the capillary. During evaporation, the particles continuously moved from the bottom to the top of the capillary, with no significant reverse convective mass transfer observed. This indicates that liquid mass transfer during capillary evaporation is predominantly unidirectional, with limited large-scale convective mass transfer, which is an important condition for non-selective evaporation.

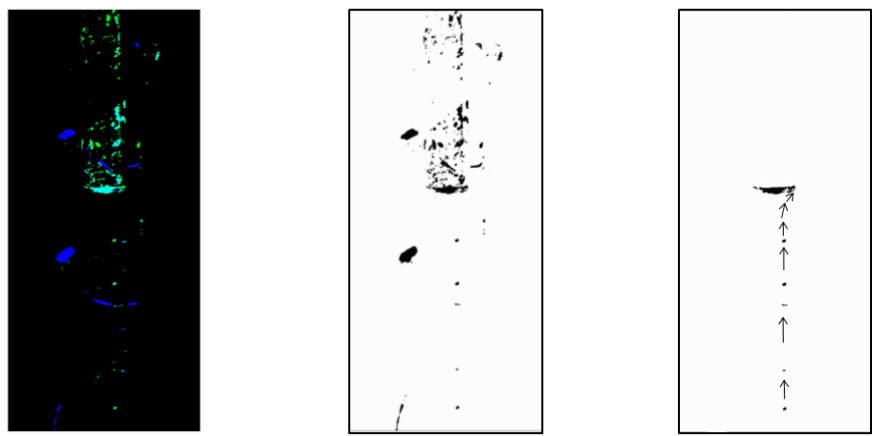

a) Original Image    b) Instantaneous Position    c) Movement Trajectories
Figure 18. Movement Trajectories of Fluorescent Particles

### 3.4.5 Microscopic Observation of Micro-Region Liquid Flow in Capillary Evaporation
To further observe micro-region liquid flow near the capillary evaporation surface, the microscopic observation setup shown in Figure 5 was used with 5 μm polystyrene microspheres as tracer particles in a 0.6 mm inner diameter capillary filled with a 1:9 methanol-water mixture. The microspheres were dispersed in 10 mL of the mixture, which was injected into the capillary, filling approximately 3/4 of its length.

At a heating temperature of 65 ℃, the movement of the tracer particles was recorded, and flow lines were drawn based on their trajectories, as shown in Figure 19. The liquid in the capillary can be divided into four regions: (1) the transport region, where liquid flows unidirectionally from the bulk liquid to the meniscus; (2) the meniscus center region, where liquid flows toward the liquid-solid interface at the meniscus edge, with minor internal recirculation; (3) the wall adsorption region, where liquid is strongly bound to the capillary wall and contributes little to evaporation; and (4) the thin film evaporation region, where evaporation primarily occurs due to high heat transfer efficiency and weak liquid-solid interactions. Local convective flow lines were observed below this region.

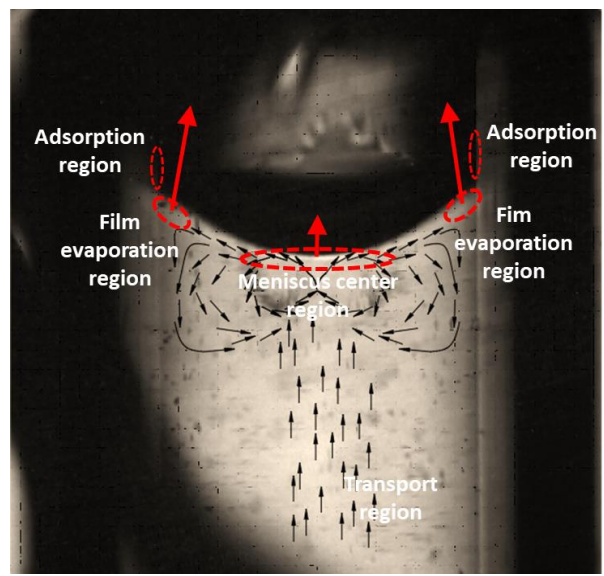

**Figure 19. Flow Trajectories of Tracer Particles in the Capillary**

The meniscus and the micro-convection region below it constitute the "evaporation layer" described in Section 3.3. Due to micro-convection, the liquid composition in the evaporation layer undergoes internal exchange, but after achieving "non-selective evaporation," it differs from the bulk liquid composition in the transport region. As shown in Figure 19, the thickness of the evaporation layer is approximately half the capillary inner diameter.

**4 Conclusions**

1) The formation of non-selective evaporation in porous medium atomizers involves three stages: selective evaporation, formation of a heterogeneous evaporation layer, and sustained non-selective evaporation. The key to this process is the formation of an evaporation layer with a composition different from that of the bulk liquid. In porous medium atomizers, the proximity of the heat source to the evaporation surface and the hindered mass transfer due to capillary action significantly contribute to the formation of the heterogeneous evaporation layer. Based on binary phase diagram analysis, when the gas-phase composition evaporating from the evaporation layer matches the liquid-phase composition entering the evaporation layer, stable non-selective evaporation is achieved. The evaporation temperature measurements, surface heating experiments, porous medium surface heating experiments, single-capillary evaporation rate measurements, and single-capillary liquid flow observations strongly support this mechanism.

2) The evaporation temperature of binary systems significantly deviates from the bubble point temperature corresponding to the bulk liquid-phase composition but closely matches the dew point temperature of the gas phase with the same composition. This indicates the existence of a relatively stable liquid-phase evaporation layer with a composition different from that of the bulk liquid. The existence of the "evaporation layer" was further validated through capillary evaporation observations.

3) Surface heating and hindered convective mass transfer play crucial roles in non-selective evaporation. Porous media simultaneously promote surface heating and increase mass transfer resistance, making them essential for non-selective evaporation in atomizers.